\documentclass[sigconf]{acmart}

\setcopyright{none}

\AtBeginDocument{%
  }

\usepackage{enumerate}

\usepackage[utf8]{inputenc}
\usepackage{tablefootnote}
\usepackage{algorithm2e}
\usepackage{graphicx}
\usepackage{xcolor}
\usepackage{multirow}
\usepackage{framed}
\usepackage{url}
\usepackage{xurl}
\usepackage{mathtools}
\usepackage{hyperref}
\usepackage{tikz}
\usepackage{amsmath}
\usepackage{pifont}
\usepackage{algpseudocode}
\usepackage[utf8]{inputenc}
\usepackage{amsthm}
\usepackage{textcomp,booktabs}
\usepackage{tabularx}
\usepackage[caption=false,font=normalsize,labelfont=sf,textfont=sf]{subfig}

\hypersetup{
    linkbordercolor = {green},
}

\usepackage{flushend}

\copyrightyear{2025}
\acmYear{2025}
\setcopyright{rightsretained}
\acmConference[PETRA '25]{The PErvasive Technologies Related to Assistive Environments}{June 25--27, 2025}{Corfu Island, Greece}
\acmBooktitle{The PErvasive Technologies Related to Assistive Environments (PETRA '25), June 25--27, 2025, Corfu Island, Greece}
\acmPrice{}
\acmDOI{10.1145/3733155.3734898}
\acmISBN{979-8-4007-1402-3/25/06}

\begin{document}

\title{The Language of Influence: Sentiment, Emotion, and Hate Speech in State Sponsored Influence Operations}

\author{Ashfaq Ali Shafin}
\email{shafinashfaqali21@gmail.com}
\affiliation{%
  \institution{Florida International University}
  \city{Miami}
  \state{Florida}
  \country{USA}
}

\author{Khandaker Mamun Ahmed}
\email{khandakermamun.ahmed@dsu.edu}
\affiliation{%
  \institution{Dakota State University}
  \city{Madison}
  \state{South Dakota}
  \country{USA}}


\begin{abstract}

State-sponsored influence operations (SIOs) have become a pervasive and complex challenge in the digital age, particularly on social media platforms where information spreads rapidly and with minimal oversight. These operations are strategically employed by nation-state actors to manipulate public opinion, exacerbate social divisions, and project geopolitical narratives, often through the dissemination of misleading or inflammatory content. Despite increasing awareness of their existence, the specific linguistic and emotional strategies employed by these campaigns remain underexplored. This study addresses this gap by conducting a comprehensive analysis of sentiment, emotional valence, and abusive language across 2 million tweets attributed to influence operations linked to China, Iran, and Russia, using Twitter's publicly released dataset of state-affiliated accounts. We identify distinct affective and rhetorical patterns that characterize each nation’s digital propaganda. Russian campaigns predominantly deploy negative sentiment and toxic language to intensify polarization and destabilize discourse. In contrast, Iranian operations blend antagonistic and supportive tones to simultaneously incite conflict and foster ideological alignment. Chinese activities emphasize positive sentiment and emotionally neutral rhetoric to promote favorable narratives and subtly influence global perceptions. These findings reveal how state actors tailor their information warfare tactics to achieve specific geopolitical objectives through differentiated content strategies.

\end{abstract}


\maketitle

\section{Introduction}

Online Social Networks (OSNs) have emerged as a notable medium for the dissemination of news, information, and public opinion in the contemporary digital landscape. Individuals across the globe increasingly rely on these platforms to express their viewpoints and engage with their social networks. Despite their numerous benefits, OSNs are frequently targeted by inauthentic entities, including malicious bots designed for automated activities and sock puppet accounts intended to misrepresent user identities, thereby exploiting inherent vulnerabilities within these platforms.

These inauthentic accounts are systematically deployed to polarize communities, propagate misleading information, and manipulate public discourse. Their strategies often involve disseminating emotionally charged content to captivate audiences and targeting dissenting voices through abusive language and coordinated harassment campaigns ~\cite{AMIK24}. The implications of such activities are particularly alarming during critical events, such as electoral processes, public health crises, and natural disasters, where coordinated networks of inauthentic accounts, frequently originating from specific geopolitical regions, are mobilized to exert undue influence on public opinion ~\cite{EPZBNT23, MD20}. The pervasive influence of these accounts not only undermines the integrity of information ecosystems but also poses significant challenges to the maintenance of democratic processes and social cohesion. Consequently, there is an urgent need for comprehensive strategies to detect, analyze, and mitigate the impact of inauthentic activities within OSNs.

Over the last few years, there has been a notable surge in attempts to sway public opinion through fraudulent social media profiles which is generally known as influence or information operations (IOs). Accounts involved in influence operations predominantly focus on spreading propaganda, false information, and hyper-partisan news to manipulate public opinion. 
Although social media platforms have made several attempts over the years to stop these influence campaign, it is not enough as these campaigns efforts become innovative~\cite{RCHS23, MBNPAPP23, ZCDSSB19}. Therefore, in order to understand and identify the language of misleading influence from social media platforms, in this paper, we investigate the content shared by the state-sponsored influence operations on X/Twitter from three distinct countries: China, Iran, and Russia. In this paper, we analyzed the sentiment, emotion, and abusive nature of the tweets shared by users (IO operators) involved in influence operation and compared the nature of content among the three IOs. We used existing tools like TweetNLP~\cite{tweet_nlp} and Perspective API~\cite{perspective_api}, specifically designed to evaluate text content from social networks.    

Our analysis reveals significant differences in the nature of influence operation content when compared across countries based on the goals of influence operations. We found that Russian IOs predominantly post negative content designed to evoke strong negative emotions, often incorporating abusive and toxic language. In contrast, IO content from China is primarily positive, focusing on promoting favorable narratives, while Iranian IO tend to adopt a more balanced tone, reflecting a mix of perspectives. Specifically, we found the Russian IOs contain more than 9 times toxic content than Chinese IOs and 3.5 times more than IOs originated from Iran.

\section{Related Work}

Social media provides people a remarkable platform to express their opinions and connect with others. In addition to sharing personal views on social media platforms, users can redistribute the contents that resonate with them. The sharing of information on social media platforms generates a massive amount of data continuously which provides a unique opportunity to analyze the data for different kinds of studies including sentiment analysis, emotion analysis, and abusive content analysis. Social media content analysis also proves to be a valuable tool for understanding public opinion on various actions and policies \cite{piedrahita2021vaccine}. For instance, governments can utilize social media data to gauge public sentiment regarding specific regulations. In 2021, researchers in the Philippines conducted an experiment to analyze public sentiment toward the COVID-19 vaccine, providing insightful information to support informed decision-making by the government \cite{villavicencio2021twitter}. Similarly, the authors in \cite{behera2021co} applied a Co-LSTM model to study consumer reviews shared on social media, showcasing the potential of advanced techniques in deriving meaningful insights from user-generated content. Social media users play a dual role, both as contributors to the spread of abusive content and as recipients who experience its impact. Unlike content understanding, a large number of studies focused extensively on detecting and identifying abusive content such as hate speech and toxic content based on machine learning and artificial neural networks~\cite{DWMW17, GS17, BGGV17, WBO18, SM22, almohaimeed2023thos}. Mathew et al.~\cite{MDGM19} found that accounts frequently posting hateful content are densely connected on social networks tend to receive significantly more reach  compared to accounts that rarely post such abusive content. Maarouf et al.~\cite{MPF24} analyzed the virality of hate speech on X/Twitter and discovered that hate speech from verified users reach significantly more users than hate speech from non-verified regular users. \footnote{X which was previously known as Twitter. Therefore, X or Twitter was used interchangeably in this paper.}

Although social media offers a remarkable opportunity for individuals to stay connected, share their personal opinions, and access the latest news, it is often exploited by influential campaigns to control public opinion and disseminate misinformation during critical societal events \cite{dash2024decoding,brangetto2016influence,keller2020coordinated,kalpokas2016influence}. In \cite{reisach2021responsibility}, authors discussed the negative impact of user profiling on social media platforms and how user profiling can contribute to disinformation and propaganda campaigns. They have also provided some suggestions for minimizing societal risks responsible actions. Another study applied machine learning approach to distinguish influence operations in social media and tested their method on publicly available Twitter data on Chinese, Russian, and Venezuelan targeting the United States \cite{alizadeh2020content}. To explore the unique challenges and opportunities presented by the interplay between social media and traditional mass media, the Russo-Ukrainian conflict serves as a pertinent case study \cite{maschmeyer2025donetsk}. Most of the times, these disinformation campaigns on social media are carried out by coordinated malicious accounts. To identify malicious accounts, authors in \cite{sharma2021identifying} proposed a AMDN-HAGE generative model and demonstrated its effectiveness on twitter data. A general network-based framework is proposed in \cite{pacheco2021uncovering} uncover the group of coordinated accounts. The proposed method constructs behavioral traces shared among the accounts to identify them. To detect automated coordination on social media, a synchronized action framework is presented in \cite{magelinski2021synchronized} that constructs and analyzes multi-view networks. In addition, several network-based methods are also proposed to unveil coordinated online behavior \cite{nizzoli2021coordinated, weber2021amplifying, weber2020s}.

\begin{table*}[h!]
    \centering
    \begin{tabular}{ccccc}
        \hline
        \textbf{Country} & \textbf{Publish Date} &\textbf{\# IO Users}& \textbf{\# IO Tweets}
        & \textbf{IO Duration}\\ 
        \hline
        China& Sep 2019&4,324& 10,241,545& Feb 2008 - Aug 2019\\
        \hline
        Iran& Oct 2018& 660 &1,122,936&Dec 2010 - Aug 2018\\ 
        \hline
        Iran& Jan 2019& 1,868&1,840,878&Jul 2017 - Mar 2018\\ 
        \hline
        Russia& Jan 2019& 361&920,761& Aug 2010 - Nov 2018\\ 
        \hline
    \end{tabular}
    \caption{Dataset Description: provides a general overview of the dataset used in this study which contains country of the data, published date, number of IO users, number of IO tweets and IO duration.}
    \label{tab:data}
\end{table*}

\section{Data}

Twitter, before the acquisition of Elon Musk and re-branding the social media as X, published state-sponsored information operations datasets during the period of 2018 and 2021. Twitter has published over 141 information operation datasets originating from 21 different countries for the research communities.

These IOs were mostly conducted by Russia, Iran, China, and Venezuela, that targeted to influence both internally in their countries and externally in other foreign countries during major events like election. These datasets contained the full account timeline history (all the tweets and retweets posted by an account) of the influence operators including their basic account information (account location, description, account creation date, etc).  

We have selected four different state sponsored influence operations datasets conducted by three countries published by Twitter between 2018 and 2019. Table~\ref{tab:data} provides an overview of the four datasets we considered in this study. The datasets contain metadata of user account information and complete user timeline. We randomly selected 500 thousand English tweets from each of the four different IOs, originating from Russia, Iran, and China, resulting in a total of 2 million tweets from 3,874 unique accounts. Due to the limited number of English tweets in the 2018 Iran IO dataset, we translated the available tweets into English. As the original dataset was substantially large, we randomly selected a sizable subset to ensure computational feasibility while preserving representativeness. This random sampling approach helps maintain the diversity of the data and does not introduce temporal misalignment, as the selection was not time-bound. All of these tweets were part of the influence operations. State-sponsored influence operations employed sophisticated manipulation tactics across social media platforms, with Iranian networks targeting Israel-related discourse through coordinated deceptive behaviors, Chinese operations utilizing thousands of accounts to amplify Chinese Communist Party (CCP) narratives and disseminate Hong Kong disinformation, and Russian campaigns conducting inauthentic amplification to promote topics related United Russia while discrediting political dissidents. These coordinated information warfare campaigns demonstrate the systematic exploitation of digital ecosystems by nation-state actors to manipulate public discourse and advance geopolitical objectives. Our decision to focus exclusively on English tweets is that foreign information operations mostly targeted counties that are native English speakers like USA and UK.

While newer datasets such as the 2021 release are available, we selected the 2019 dataset because it is substantially larger and includes operations from all countries of interest. In contrast, the 2021 dataset is smaller and does not include all relevant countries, limiting its suitability for our analysis.

\section{Methodology}
Our method starts the analysis of the dataset by pre-processing it first. To analyze the text content of our dataset, we preprocessed the text data. We remove unnecessary characters such as new lines (`\textbackslash n'), carriage return (`\textbackslash r') and convert HTML-encoded characters to the original form (e.g., `\&amp' to `\&'). Further, we replaced all the URLs and user mentions from the tweets with `[URL]' and `[user]'. 

In this paper, we applied TweetNLP's ~\cite{tweet_nlp}, a content understanding tool for social network. Specifically, we utilized sentiment, emotion, hate-speech, and offensive language detection models to identify undesirable content shared by influence operatives originated from three distinct countries. The sentiment of a tweet was classified into three distinct classes: negative, neutral, and positive, while the emotion of a tweet was identified into four categories: anger, joy, optimism, and sadness. Both hate-speech and offensive language models were based on binary classes where positive means the tweet text contains hate-speech/offensive language and negative indicates non-hate-speech/non-offensive language. 

Moreover, we used Perspective API~\cite{perspective_api}, a toxicity detection model developed by Google to identify toxic content from the information operations. Perspective API provided six distinct classes for each of the tweets: toxic, severe toxic, profanity, identity attack, insult, and threat. Perspective API model is based on multi-label classification meaning one tweet can belong to multiple output classes. We detailed our methodology in Algorithm~\ref{algo:io}. The input for the algorithm is the raw tweets from the Twitter SIO dataset and the output of of the algorithm is the classification of the content among sentiment, emotion, hate, offensive and toxicity. 

\begin{algorithm}
\caption{Social media content analysis algorithm.}
\label{algo:io}
\SetKwInOut{Input}{Input}
\SetKwInOut{Output}{Output}

\Input{IO tweet dataset $D = \{t_1, t_2, \ldots, t_n\}$}
\Output{Analyzed dataset $A = \{a_1, a_2, \ldots, a_n\}$ with classification result of content.}
\BlankLine
\emph{Initialization:}\\
Initializing empty sets for preprocessed tweets and final combined result:
\[
D' = \emptyset, C = \emptyset
\]
\BlankLine
\emph{Preprocessing Step:}\\
\ForEach{tweet $t_i \in D$}{
    Remove unnecessary characters:
    \[
    t'_i = t_i \setminus \{\text{\textbackslash n,...,\textbackslash r}\}
    \]
    Decode HTML-encoded characters:
    \[
    t'_i = f_{\text{decode}}(t'_i) \quad \text{(e.g., \&amp; $\to$ \&)}
    \]
    Replace URLs and mentions:
    \[
    t'_i = f_{\text{url}}(t'_i), \quad t'_i = f_{\text{mention}}(t'_i)
    \]
    Store preprocessed tweet:
    \[
    D' = D' \cup \{t'_i\}
    \]
}

\BlankLine
\emph{Analysis Step:}\\

\ForEach{tweet $i \in D'$}{
    \[
    S_i = f_{\text{sentimentTweetNLP}}(i)
    \]
    \[
    E_i = f_{\text{emotionTweetNLP}}(i)
    \]
    \[
    H_i = f_{\text{hateTweetNLP}}(i)
    \]
    \[
    O_i = f_{\text{offensiveTweetNLP}}(i)
    \]
    \[
    T_i = f_{\text{toxicityPerspective}}(i)
    \]
    Combine analysis results:
    \[
    c_i = \{S_i, E_i, H_i, O_i, T_i\}
    \]
    \[
     C = C \cup \{c_i\}
    \]
}

\BlankLine
\Return Combined dataset:
\[
C = \{c_1, c_2, \ldots, c_n\}
\]

\end{algorithm}

\section{Result}

This section presents a comprehensive analysis of two million tweets distributed by the Chinease, Iranian, and Russian influence operators. We explored sentiment, emotion, hate speech, offensive language, and overall content toxicity to discover patterns and strategies employed by IO operators. The results reveal distinct behavioral and linguistic trends across the three countries, shedding light on their targeted manipulation tactics.

\begin{table*}[h!]
    \centering
    \begin{tabular}{ccccccccc}
        \hline
        \textbf{Dataset} & \textbf{\# Tweets} & \multicolumn{3}{c}{\textbf{Sentiment}} & \multicolumn{4}{c}{\textbf{Emotion}} \\ 
        \cmidrule(lr){3-5} \cmidrule(lr){6-9}
         &  & \textbf{Negative} & \textbf{Neutral} & \textbf{Positive} & \textbf{Anger} & \textbf{Joy} & \textbf{Optimism} & \textbf{Sadness}\\ 
        \hline
        China-19 & 500,000& 31,026 & 295,005 & 173,969 &38,645 & 340,315& 81,676& 39,364\\ 
        \hline
        Iran-18 & 500,000 & 101,289 & 192,294 & 206,417 & 128,121 & 184,365 &102,122&85,392\\
        \hline
        Iran-19 & 500,000 & 144,666 & 245,487 & 109,847 & 200,567 & 112,944 &134,477&52,012\\ 
        \hline
        Russia-19 & 500,000 & 205,882 & 224,930 & 69,188 & 286,871 & 87,239& 70,947&54,943\\ 
        \hline
    \end{tabular}
    \caption{Sentiment and emotion results of the SIO content. }
    \label{tab:sentiment}
\end{table*}

\begin{table*}[h!]
    \centering
    \begin{tabular}{cccccc}
        \hline
        \textbf{Dataset} & \textbf{\# Tweets} & \multicolumn{2}{c}{\textbf{Hate Speech}} & \multicolumn{2}{c}{\textbf{Offensive}} \\ 
        \cmidrule(lr){3-4} \cmidrule(lr){5-6}
         &  & \textbf{Positive} & \textbf{Negative} & \textbf{Positive} & \textbf{Negative}\\ 
        \hline
        China-19 & 500,000& 8,668& 491,332& 8,200&491,800\\ 
        \hline
        Iran-18 & 500,000 & 36,175& 463,825& 20,199& 479,801\\
        \hline
        Iran-19 & 500,000 & 40,824& 459,176& 21,776& 478,224\\ 
        \hline
        Russia-19 & 500,000 & 89,220& 410,780& 54,401 & 445,599\\ 
        \hline
    \end{tabular}
    \caption{TweetNLP Hate Speech and Offensive Result}
    \label{tab:hatespeech}
\end{table*}

\begin{table*}[h!]
    \centering
    \begin{tabular}{ccccccccc}
        \hline
        \textbf{Dataset} & \textbf{\# Tweets} & \textbf{Toxic} & \textbf{Severe Toxic} & \textbf{Profanity} & \textbf{Identity Attack} & \textbf{Insult} & \textbf{Threat}&\textbf{Overall}\\  
        \hline
        China-19 & 500,000& 4,511& 187& 3,970&219&2,034&535& 5,863\\ 
        \hline
        Iran-18 & 500,000 & 6,715 & 513 & 2,092 & 4,951 & 6,102 & 3,191 & 13,347 \\
        \hline
        Iran-19 & 500,000 & 7,760& 640& 2,341& 5,356&7,252&3,401& 15,048\\ 
        \hline
        Russia-19 & 500,000 & 37,128& 7,392& 12,613& 27,185&26,964&5,505& 53,464\\ 
        \hline
    \end{tabular}
    \caption{Toxicity results from Perspective API.}
    \label{tab:Toxicity}
\end{table*}

\noindent
{\bf Sentiment Analysis}.
The sentiment analysis, summarized in Table \ref{tab:sentiment}, highlights differences in the tone and intent of the tweets across the three countries. Tweets from Chinese IOs were predominantly neutral (59.0\%, 295,005 tweets), suggesting a focus on information dissemination with minimal emotional engagement. Positive sentiment was also notable (34.8\%, 173,969 tweets) likely reflecting a strategic attempt to project optimism and reinforce favorable narratives whereas negative sentiment was comparatively low (6.2\%, 31,026 tweets), indicating limited deployment of confrontational or divisive rhetoric. Compared to 2018, Iranian IO tweets published in 2019 displayed a more confrontational tone, with a substantial proportion classified as negative (28.9\%, 144,666 tweets). Neutral sentiment remained dominant (49.1\%, 245,487 tweets), while positive sentiment (21.9\%, 109,847 tweets) was the lowest among the three categories. This distribution aligns with Iran’s likely use of IOs to spread discontent, counter opposition, and amplify divisive narratives. Russian IO tweets stood out for their highly negative sentiment (41.2\%, 205,882 tweets), along with the neutral (44.9\%, 224,930 tweets) and positive (13.8\%, 69,188 tweets) tones. The predominance of negativity suggests a deliberate effort to foster division, amplify discontent, and destabilize target audiences. The analysis of sentiment revealed distinct strategies across the countries: China emphasized neutrality and positivity to promote favorable narratives, Iran leaned towards negativity to provoke hostility and division, while Russia predominantly utilized negative sentiment to confront the target.

\noindent
{\bf Emotion Analysis}.
The emotion analysis is summarized in Table \ref{tab:sentiment} that provides insights into the rhetorical strategies employed by IO actors from three different countries. Tweets from Chinese operators were dominated by positive emotions, particularly joy (68.1\%, 340,315 tweets) and optimism (16.3\%, 81,676 tweets). Anger (7.7\%, 38,645 tweets) and sadness (7.9\%, 39,364 tweets) were considerably less frequent, reflecting a tendency to emphasize favorable narratives and promote unity. Tweets from Iranian IOs displayed a relatively balanced mix of negative and positive emotions. Anger was the defining emotion in Iranian 2019 IO tweets, accounting for 40.1\% (200,567 tweets), highlighting their use of emotionally charged content to provoke hostility or resentment and sadness (10.4\%, 52,012 tweets) played a smaller but significant role, potentially used to evoke sympathy or amplify grievances. Compared to 2018 Iranian IOs, 2019 IOs contained more negative emotions than positive emotions in the tweets.
In contrast, positive emotions such as joy (22.6\%, 112,944 tweets) and optimism (26.9\%, 134,477 tweets) were also notable, perhaps reflecting a dual strategy of rallying support while vilifying adversaries. Similar to sentiment analysis, Russian tweets exhibited the highest levels of anger (57.4\%, 286,871 tweets), underscoring their focus on polarizing audiences and amplifying frustration. Joy (17.4\%, 87,239 tweets) and optimism (14.2\%, 70,947 tweets) were less common, while sadness (11.0\%, 54,943 tweets) was the least portrayed emotion. It is evident from the result, while China tried to evoke positive emotions from their targets, Russian operators were mostly divisive in their content to generate negative emotions. Iranian IOs designed their content in a more balanced emotional settings to their content.

\begin{figure*}[t]
\centering
\subfloat[]{\includegraphics[width=0.32\textwidth]{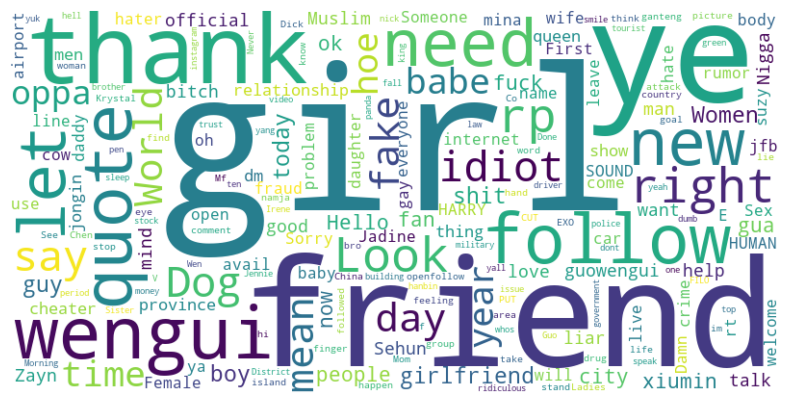} \label{fig:china}}
\hfill
\subfloat[]{\includegraphics[width=0.32\textwidth]{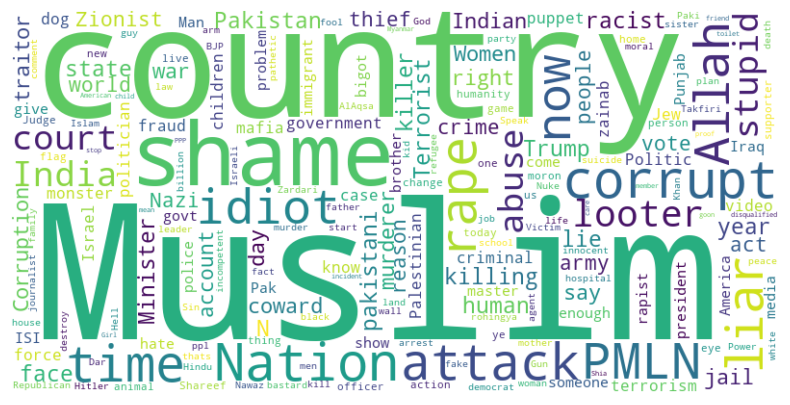} \label{fig:iran}}
\hfill
\subfloat[]{\includegraphics[width=0.32\textwidth]{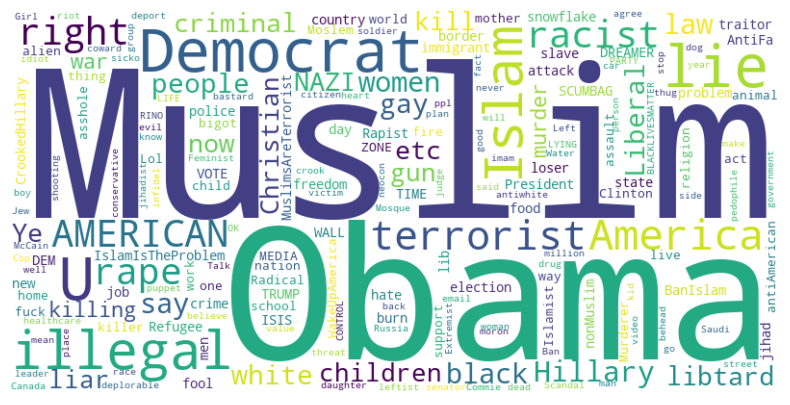} \label{fig:russia}}
\hfill
\vspace{-8pt}

\caption{Word Cloud generated for abusive speech from (a) Chinese, (b) Iranian, and (c) Russian IOs. Abusive speech considered tweets containing hate speech, offensive langue, or toxic content.}
\label{word_cloud}
\vspace{-8pt}
\end{figure*}
\noindent
{\bf Hate Speech and Offensive Language Analysis.}
The hate speech and offensive language detected by TweetNLP model is shown in Table \ref{tab:hatespeech}, reveals varied approaches to the use of harmful rhetoric by the IO operators. Among Chinese IO tweets, only a small proportion was classified as hate speech (1.73\%, 8,668 tweets) or offensive (1.64\%, 8,200 tweets). This suggests a preference for content that avoids inflammatory language, possibly to comply with platform policies. Russian IOs demonstrated the highest levels of hate speech (17.84\%, 89,220 tweets) and offensive-positive (10.88\%, 54,401 tweets) content, reflecting their dependency on divisive and incendiary rhetoric. Comparative analysis of Iranian IOs between 2018 and 2019 reveals escalation in abusive contnet. In 2019, hate speech comprised 8.16\% of total tweets (40,824 tweets), while offensive-positive content accounted for 4.3\% (21,776 tweets), representing substantial increases from the previous year. Specifically, we observed more than 12\% increase in hate speech and 7\% rise in offensive language, indicating a strategic shift toward more confrontational messaging tactics designed to amplify divisive content and target specific audiences.

\noindent
{\bf Toxicity Analysis}.
We analyzed six different toxicity categories (Toxic, Severe Toxic, Profanity, Identity Attack, Insult, and Threat) and represented the overall result of the identified tweets in Table \ref{tab:Toxicity}. All the IOs exhibited more toxic tweets than any other categories. The overall score of toxicity across the Chinese IO tweets of 5863 (1.17\%), with specific categories as severe toxicity (187) and threats (535) making it clear that very few overtly harmful strategies are employed. Tweets from Iranian IOs show more general toxicity, with an overall total of 15,048 (3.01\%). Identity attacks (5,356) and insults (7,252) are particularly marked, indicating that they would focus on specific people or groups. The highest toxicity measure demonstrated by the Russian IOs, with an overall score of 53,464 tweets (10.69\%). Toxicity (37,128 tweets), identity attacks (27,185 tweets) and insults (26,964 tweets) clearly characterize an aggressive and polarizing narrative strategy.

Based on the Perspective API toxicity analysis, Iranian Information Operations demonstrated a substantial escalation in toxic content between 2018 and 2019. Overall toxicity increased by 12.7\% (from 13,347 to 15,048 tweets), with particularly pronounced increases in severe toxicity (24.8\% increase, from 513 to 640 tweets) and insulting language (18.8\% increase, from 6,102 to 7,252 tweets). General toxic content rose by 15.6\% (6,715 to 7,760 tweets), while identity attacks increased by 8.2\% (4,951 to 5,356 tweets). Profanity and threats showed more moderate increases of 11.9\% and 6.6\%, respectively. This systematic increase across all toxicity categories suggests a deliberate strategic shift toward more aggressive and harmful discourse, reflecting an intensification of Iranian IO tactics designed to maximize social discord and polarization in target populations.

Figure ~\ref{word_cloud} shows the word cloud, a visual representation of the most common term used in abusive content considering hate speech, offensive speech, and toxic content for the three different IOs in 2019. The size of the words are proportional to its frequency of occurrence. China targets dissidents like Guo Wengui, an exiled Chinese businessman to discredit him. Chinese IOs also use positive terms such as ``thank", ``friend" to build trust among the foreign social media users. Iran emphasizes more on religious and nationalistic themes, with words like ``Muslim", ``nation," and ``shame", targeting adversaries and presenting itself as a leader among the Muslim nations. Russia exploits political and cultural divisions, especially in the U.S., using terms like ``Obama", ``Democrat", and ``gun" to polarize and erode democratic trust. These strategies reflect diverse information tactics aligned with each state’s objectives and interests.

\section{Discussion and Limitations}
In this article, we have focused on content analysis of three distinct influence operations conducted by China, Iran, and Russia on social media X/Twitter. From our analysis, it is evident that IOs from different countries differ in their content in terms of their sentiment, emotions, and toxicity. Russian IOs contain more negative sentiment with negative emotions like anger and sadness compared to the Chinese and Iranian IOs. Chinese IOs propagate more positivity trying to evoke positive emotions in their content. The variation in content tone may be attributed to the differing objectives of IOs across countries. For instance, while Russia is responsible for disrupting major events like elections, whereas China is more likely to conduct their information operations to convince target audience about the positive impact of China and their leading political party Chinese Communist Party (CCP). 

Further, we can see that information operations contain varying levels of abusive content and toxicity. IOs originating from Russia accumulate almost twice the amount of hate speech and offensive tweets compared to the IOs distributed by both Iran and China. In this paper, we only considered tweets in English for analysis. Due to the scarcity of different models for content analysis, we could not examine the content in different languages. However, our research with limited data still provides valuable insight regarding the domain of influence operations unlike previous experiments that did not focus on content analysis in situations of state sponsored influence operations.

Despite the valuable insights provided by our analysis, this study has its own limitations. Our study did not perform a comprehensive error analysis, which limits our ability to systematically identify misclassifications or biases in the model’s outputs. Future work should include a detailed error analysis to better understand the model’s failure modes in different linguistic and contextual settings. Moreover, the ML model used in this paper was primarily trained on Western discourse patterns, which limits the model's ability to accurately capture abusive content across different cultural contexts. Therefore, future work can include the cultural influence of model's performance. In addition, we have randomly selected a subset of the original dataset due to the enormous size of the dataset. Although the size of the subset is substantial, study on the entire dataset can provide more coherent temporal alignment.

\section{Conclusion} State controlled influence operations leverage fast information diffusion functionality of social media to manipulate and deceive authentic users to change their perceptions. Our study reveals IO campaigns initiated from different countries show distinct characteristics in their narratives in terms of sentiment, emotions, and toxicity. This indicates that influence operators frequently leverage abusive content to entice their followers to achieve their ultimate goals and vary their abusive content strategy based on the need of the operations. 

\bibliographystyle{ACM-Reference-Format}
\bibliography{hatespeech, IO}

\end{document}